\documentstyle[12pt]{article}
\begin{document}
\begin{titlepage}
\vspace{2.5cm}
\begin{centering}
{\LARGE{\bf Geodesics around line}}\\
\bigskip
{\LARGE{\bf defects in elastic solids }}\\
\bigskip\bigskip
A. de Padua, Fernando Parisio-Filho and Fernando Moraes\\
{\em Departamento de F\'{\i}sica\\
Universidade Federal de Pernambuco\\
50670-901 Recife, PE, Brazil}\\
\vspace{1cm}

\end{centering}
\vspace{1.5cm}

\begin{abstract}
Topological defects in solids, usually described by complicated boundary conditions in elastic theory, may be described more simply as sources of a gravity-like deformation field in the geometric approach of Katanaev and Volovich.  This way, the deformation field is described by a non-Euclidean metric that incorporates the boundary conditions imposed by the defects. A possible way of gaining some insight into the motion of particles in a medium with topological defects (e.g., electrons in a dislocated metal) is to look at the geodesics of the medium around the defect. In this work, we find the exact solution for the geodesic equation for an elastic medium with a generic line defect, the dispiration, that can either be a screw dislocation or a wedge disclination for particular choices of its parameters. 

\end{abstract}

\hspace{0.8 cm}
\end{titlepage}
\def\carre{\vbox{\hrule\hbox{\vrule\kern 3pt
\vbox{\kern 3pt\kern 3pt}\kern 3pt\vrule}\hrule}}

\baselineskip = 18pt

Topological defects play an important role in determining the physical properties of real materials. Mechanical and electronic properties as well as phase transitions are strongly affected by their presence. A topological defect consists in a core region, characterized by the absence of order, and a smooth far field region. In the continuum approximation the core is shrunk to a singularity as will be seen below. Topological defects, although formed during phase transitions involving symmetry breaking, can be
conceptually generated by a ``cut and glue" process, known in the literature as the Volterra process~\cite{Kle1}. This process gives a unifying view of the topological line defects. That is, take a cylinder of a continuous elastic material and make a radial cut in it, from its axis out. Displacement of the surfaces of the cut with respect to each other and subsequent glueing  will generate a line defect whose core coincides with the axis. Considering cylindrical coordinates, if the displacement is:

(i) along the $z$-direction  a  screw dislocation is formed.

(ii) along the $\rho$-direction, an edge dislocation is formed.

(iii) along the $\theta$ direction, which implies the addition or removal of a wedge of material, leads to a disclination.

(iv) a combination of both (i) and (iii) it produces a dispiration.
 
It is clear then, that the core of such line defects are associated to geometric singularities. Here we come close to gravity theory, where geometric singularities are sources of gravitational field which, on the other hand, is described by a deformation of the space-time from the flat Minkowsky geometry.

It is well known that elastic solids with topological defects can be described by Riemann-Cartan geometry~\cite{Kro}. Recently, Katanaev and Volovich~\cite{Kat} have shown the equivalence between three-dimensional gravity with torsion and the theory of defects in solids. The defect acts as a source of a ``gravitational" distortion field. The metric describing the medium surrounding the defect is then a solution to the three-dimensional Einstein-Cartan equation. Based in this formalism, the non-relativistic quantum mechanics of electrons or holes around disclinations~\cite{Fur1,Fur2} and some properties of the quantum electromagnetic field also around disclinations~\cite{Mor1,Mor2} have been studied. In a recent letter~\cite{Mor3} one of us presented a study of the geodesics around an edge dislocation. In this work we extend this study to disclinations and screw dislocations. Disclinations are common in lower dimensional systems like graphite~\cite{xxx} or liquid crystals~\cite{Kle1} and an important ingredie
nt in the making of geometrically frustrated amorphous solids~\cite{Kle2}. They do not appear in bulk crystals, though, due to the very high elastic energy involved in their formation. Dislocations do appear in ordinary 3D crystals, but the asymmetry they provoke in the medium makes it difficult to study dynamical properties of objects moving in their midst. 

By presenting the geodesics around the line defects, this work contributes to the visualization of the deformation of the medium around them and hopefully, will provide a means of obtaining some insight into the motion of particles and the geometrical optics in defected media. Previous work on geodesic motion around defects has been done by de Sousa Gerbert and Jackiw~\cite{Ger} who studied the motion around a spinning conical singularity in (2+1) dimensions and by Gal'tsov and Letelier~\cite{Gal} who studied the relativistic motion around a spinning string and around a cosmic dispiration.

Topological defects in the framework of traditional elasticity theory are described by complicated boundary conditions. One of the advantages of the gravitational approach is the simplicity it describes the defects. In this approach, the elastic solid without defects is assumed to be a continuous, infinite medium whose undeformed state is characterized by a flat Euclidean metric $\delta_{ij}$. Ordinary elasticity theory~\cite{Lan} attaches a Cartesian reference frame $x^{i}$ to the undistorted medium and describes its deformations locally by the displacement vector field $u^{i}(x)$, which is a smooth function. This way, after the deformation, the point $x^{i}$  will have the coordinates $y^{i}=x^{i}+u^{i}(x)$ in the initial Cartesian frame. The initial Euclidean metric is then transformed into the metric
\begin{equation}
g_{ij}(x)=\frac{\partial x^{k}}{\partial y^{i}}\frac{\partial x^{i}}{\partial y^{j}}\delta_{kl}.
\end{equation}

In Riemannian geometry without torsion the curvature tensor is given by~\cite{Mis}
\begin{equation}
R_{ijk}^{l}(x)=\partial_{i}\Gamma_{jk}^{i} + \Gamma_{im}^{l}\Gamma_{jk}^{m} - \partial_{j}\Gamma_{ik}^{j} + \Gamma_{jm}^{l}\Gamma_{ik}^{m},
\end{equation}
where the connection, uniquely defined by the metric, is
\begin{equation}
\Gamma_{ijk}=\Gamma_{ij}^{l}g_{lk}=\frac{1}{2}(\partial_{i}g_{jk}+ \partial_{j}g_{ik}- \partial_{k}g_{ij}).
\end{equation}
It is obvious that the curvature tensor of the undeformed metric is zero. It is also zero in the deformed medium  since the transformation (1) is a diffeomorphism\footnote{An infinitely differentiable 1-1 map between two infinitely differentiable manifolds (in this case both are $R^3$).} and it transforms covariantly under diffeomorphisms. In other words, even though $R_{ijk}^{l}(x)\equiv 0$, the deformation is present and described by the metric (1). In the language of general relativity, we have here gravitation in the absence sources. 

Topological defects will act as sources of deformation just as mass distributions do in the case of gravity. But the defects give rise not only to curvature but also to torsion. In fact, dislocations, which are associated to broken translational symmetry, are line sources of torsion whereas disclinations, associated to broken rotational symmetry, are line sources of curvature. This is clear from the relations for the Burgers~\cite{Lan} $b^{i}$ and the Frank~\cite{Kle1} $\Omega^{ij}$ vectors 
\begin{equation}
b^{i}=-\oint_{C}dx^{m}\partial_{m}u^{i}(x)
\end{equation}
and
\begin{equation}
\Omega^{ij}=\oint_{C}dx^{m}\partial_{m}\omega^{ij},\,\,\,\,\omega^{ij}\in SO(3),
\end{equation}
where $C$ is a closed curve around the defect line. In the framework of the gravitational theory of defects the above equations become respectively
\begin{equation}
b^{i}=-\int\int_{S}dx^{m}\wedge dx^{n}T_{mn}^{i}
\end{equation}
and
\begin{equation}
\Omega^{ij}=\int\int_{S}dx^{m}\wedge dx^{n}R_{mn}^{ij},
\end{equation}
where $S$ is a surface with boundary $C$ and perpendicular to the defect line. Generalizing to a continuous distribution of defects, the torsion tensor is interpreted as the surface density of the Burgers vector field, while the curvature is the surface density of the Frank vector field. For the case of a single line defect the corresponding torsion or curvature tensor will be a $\delta$-function.

Metrics describing a single edge dislocation or a single wedge disclination have been found by Katanaev and Volovich~\cite{Kat}. Tod~\cite{Tod} found the metric that describes a single wedge dispiration~\cite{Har}, a defect that includes both a screw dislocation and a wedge disclination as particular cases. Next, we solve the geodesic equations for these metrics and sketch a few representative trajectories. 

The geodesics in a defected medium may be obtained by requiring that the distance between two given points  be a minimum. That is, if $\gamma$ is a curve parametrized by $t$ joining the points at $t=t_0$ and at $t=t_1$, the metric
\begin{equation}
ds^2=g_{ij}dx^{i}dx^{j} 
\end{equation}
leads to
\begin{equation}
s[\gamma]=\int_{t_0}^{t_1} (g_{ij}\dot{x}^{i} \dot{x}^{j})^{1/2}dt.
\end{equation}
Requiring $s[\gamma]$ to be minimum, determines the path $\gamma$. The geodesic $\gamma$ is then a solution to the equation 
\begin{equation}
\frac{d^{2}x^{i}}{dt^2}+\Gamma^{i}_{jk}\frac{dx^j}{dt}\frac{dx^k}{dt}=0,
\end{equation}
where the connection coefficients or Christoffel symbols of the second kind, $\Gamma^{i}_{jk}$, are given by equation (3) and $g_{ik}g^{kj}=\delta ^j _i$. Equation (10) is essentially Newton's law of motion for a free particle in a non-Euclidean space.

Dispirations are generated by simultaneous translation and rotation break of symmetry. The wedge dispiration can be better understood if one looks first at the related defects wedge disclination and screw dislocation. The disclination is made by the insertion or removal of a wedge of material. This way, if one goes around the axis of the defect (assumed hereafter to be the $z$ axis), the total angle described is no longer $2\pi$ but some $2\pi\alpha$. If $\alpha>1$ extra material was added to the medium and if $\alpha<1$ material was removed from the medium.
In fact, this is obvious from the metric found by Katanaev and Volovich~\cite{Kat} for this defect, 
\begin{equation}
ds^{2}=dz^{2}+dr^{2}+\alpha^{2}r^{2}d\theta^{2},
\end{equation}
in cylindrical coordinates. It so happens that this metric is just a spatial section of the cosmic string metric~\cite{Vil}
\begin{equation}
ds^{2}=-c^{2}dt^{2}+dz^{2}+dr^{2}+\alpha^{2}r^{2}d\theta^{2}.
\end{equation}

Defects in cosmology have long been related to defects in solids~\cite{Hol}. It was with this in mind that Tod~\cite{Tod} studied the metric\footnote{Here we have intentionally modified the notation of reference~\cite{Tod} in order to be consistent with our own.}
\begin{equation}
ds^{2}=-c^{2}(dt+\gamma d\theta)^{2}+dr^{2}+\alpha^{2}r^{2}d\theta^{2}+(dz+\beta d\theta)^{2}.
\end{equation}
Notice that $\alpha$ describes a wedge disclination whereas $\beta$ couples the motion along the $z$-axis to the angular coordinate $\theta$. This is what is expected from a screw dislocation along the $z$-axis since, as described in the introduction, the screw dislocation is formed when after a radial cut of a cylinder, the surfaces of the cut are moved with respect to each other along the $z$-direction. That is, going around the $z$-axis one also climbs up or down (depending on the chirality of the screw) along $z$. The distance climbed after a complete loop should equal the Burgers vector. This way we may identify $2\pi\beta$ with $b^{z}$. A space-time version of the screw dislocation is described by $\gamma$. The space section of Tod's metric gives us then the wedge dispiration in the general case, the screw dislocation when $\alpha=1$, and the wedge disclination when $\beta=0$. 

In order to look for the solutions to the geodesic equation (10) the first step is to calculate the Christoffel symbols of second kind $\Gamma ^{i} _{jk}$. Symmetry reduces the number of nonzero Christoffel symbols for the dispiration metric  to only five:
\begin{equation}
 \begin{array}{l}
  \Gamma ^z _{\theta r} =\Gamma ^z _{r \theta} =\frac{-\beta}{r},\\
  \Gamma ^r _{\theta \theta} = -\alpha ^2 r,\\
  \Gamma ^\theta _{r \theta} =\Gamma ^\theta _{\theta r} =\frac{1}{r}.
 \end{array}
\end{equation}
Now, we substitute these results into the geodesic equation and get the following coupled differential equations:
\begin{equation}
\frac{d^{2}z}{dt^2} - \frac{2\beta}{r}\frac{dr}{dt}\frac{d\theta}{dt}=0 ,
\end{equation}
\begin{equation}
\frac{d^{2}r}{dt^2} - \alpha ^2 r \left(\frac{d\theta}{dt}\right)^2=0 ,
\end{equation}
and
\begin{equation}
\frac{d^{2}\theta}{dt^2} + \frac{2}{r}\frac{dr}{dt}\frac{d\theta }{dt}=0.
\end{equation}

Substituting equation (17) into (15) we get $\frac{d^{2}z}{dt^2}+ \beta\frac{d^{2}\theta}{dt^2}$ which is easily integrated to 
\begin{equation}
\frac{dz}{dt}+\beta \frac{d\theta}{dt}=A,
\end{equation}
where $A$ is an integration constant. This equation expresses the conservation of the combined momentum $\dot{z}+\beta\dot{\theta}$, a peculiarity of the space we are dealing to which couples $z$ to $\theta$. Further integration gives
\begin{equation}
z + \beta\theta = At + B,
\end{equation}
where $B$ is another integration constant. 
With simple manipulations equations (16) and (17) lead to
\begin{equation}
\alpha ^2 r^2 \frac{d\theta }{dt} = C,
\end{equation}
and 
\begin{equation}
\frac{1}{2}\left(\frac{dr}{dt}\right)^{2}+\frac{C^2}{2\alpha^{2}r^{2}}=E,
\end{equation}
where $C$ and $E$ are integration constants. Inspection of these equations reveals, respectively, the conservation of angular momentum along the $z$ direction and the conservation of energy.

The second order equations (16) and (17) are then replaced by the first order equations (20) and (21). Now, (21) can be easily solved resulting in 
\begin{equation}
r(t)=\sqrt{\frac{C^2}{E\alpha^2}+2E(t+D)^2},
\end{equation} 
where D is another integration constant. Now, substituting $r(t)$ into (20) we get 
\begin{equation}
\frac{d\theta}{dt} = \frac{2EC}{C^{2} + 4E^{2}\alpha ^2 (t + D)^2},
\end{equation}
which when integrated gives
\begin{equation}
\theta (t)= \frac{1}{\alpha} \arctan\left(\frac{2E\alpha (t + D)}{C}\right)
+ \frac{F}{\alpha},
\end{equation}
where $F$ is an integration constant. Now, inserting this result back into equation (19) we find
\begin{equation}
z(t)=At-\frac{\beta}{\alpha} \arctan\left(\frac{2E\alpha (t + D)}{C}\right)
- \frac{\beta F}{\alpha}+B.
\end{equation}

The geodesics around a single dispiration along the $z$-axis are then completely determined by the parametric equations (22), (24) and (25) for  $\alpha\neq 1$ and  $\beta\neq 0$. When $\alpha=1$,  $\beta\neq 0$ we have geodesics around a screw dislocation and when  $\alpha\neq 1$, $\beta=0$  geodesics around a wedge disclination. In Fig. 1 it is shown a  geodesic in a medium with a disclination with $\alpha$ deliberately small (1/10) in order to make evident the winding behavior of trajectories  around a deficit angle defect. In Fig. 2  the Burgers vector is turned on, transforming the defect into a dispiration. As  consequence, there is a stretching of the curve along the defect direction. Fig. 3 shows a geodesic around an excess angle disclination ($\alpha >1$) in a $z=$constant surface. The value chosen here for $\alpha$, 7/5, corresponds to a disclination defined by a seven-fold Carbon ring in the otherwise hexagonal lattice of graphite~\cite{xxx}. In Fig. 4 it is shown a geodesic asymptotically paralle
l to the $xy$-plane in a screw-dislocated medium. The effect of the dislocation is an evident lift of the geodesic from the plane.

In this letter we extend a previous study of geodesics around an edge dislocation to other line defects in solids: dispiration, screw dislocation and wedge disclination.  The simplicity of the calculation presented here demonstrates the advantage of the gravitational description of defects as compared to traditional elastic theory, where the defects are described by complicated boundary conditions. In the approach used here, the boundary conditions are expressed by a non-Euclidean metric such that the problem of finding the geodesics reduces to the study of the motion of a classical free particle in a locally curved or torsioned background space. The geodesics give a pictorial description of the deformation of the medium due to the defect. They describe, for example, the motion of classical electrons in a metallic lattice strained by topological defects. 
 
\noindent
{\bf Acknowledgments} This work was partially supported by CNPq and FACEPE. We are indebted to Jo\~ao Farias for helping in the preparation of the figures and to Claudio Furtado and M\'ario Oliveira for many important suggestions.

\newpage

\noindent
{\Large {\bf Figure Captions}}
\\
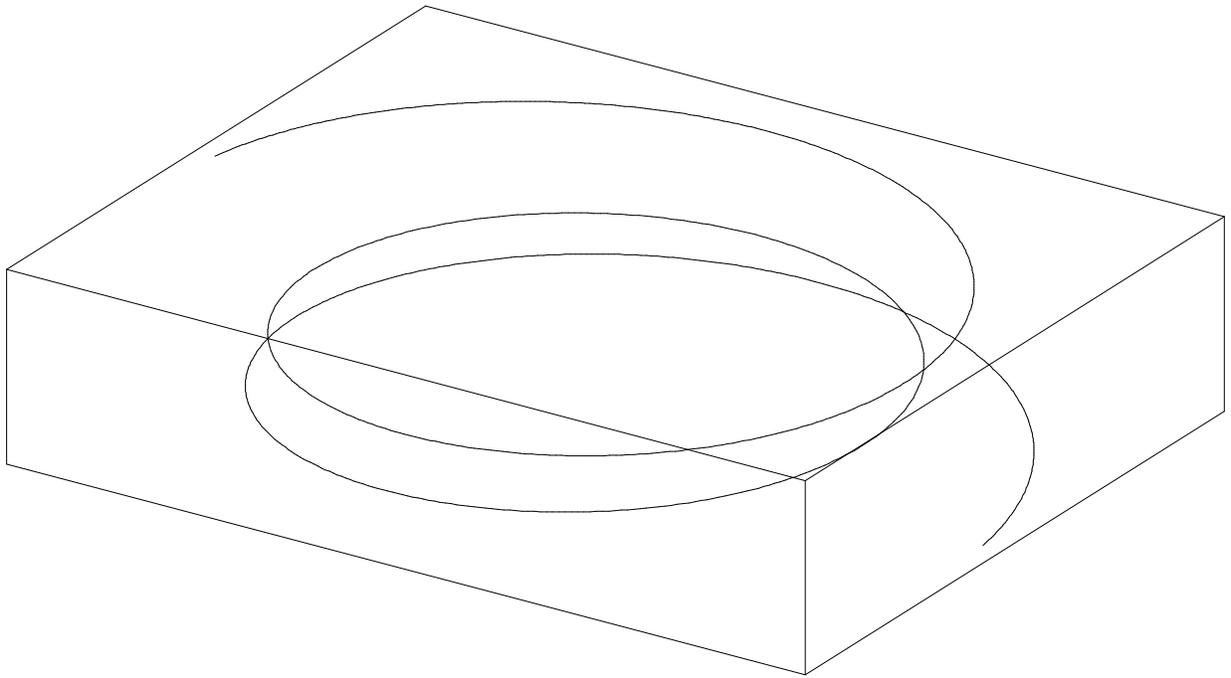
Fig. 1: Geodesic around disclinated medium, $\alpha=1/10$, $\beta=0$.
\\
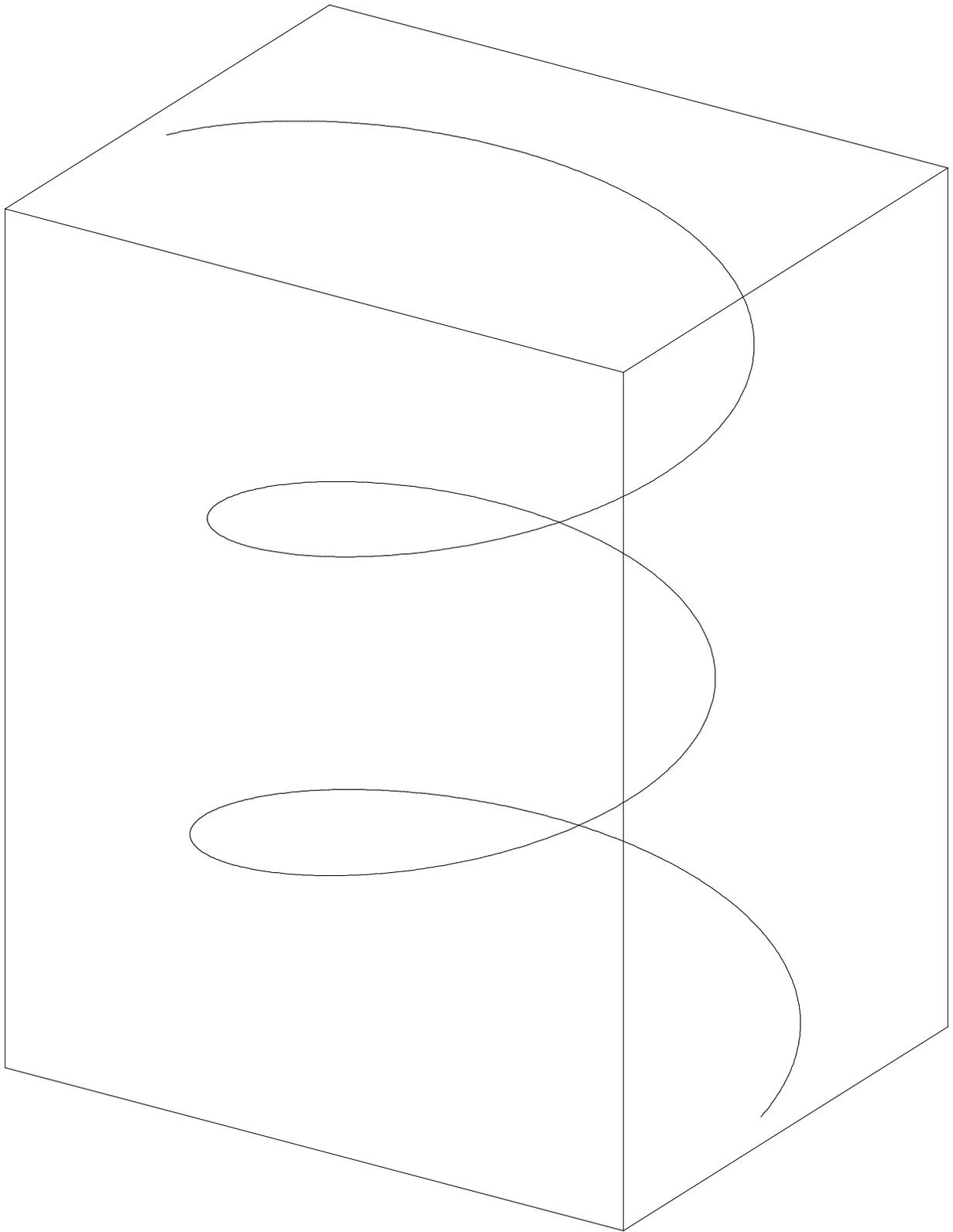
Fig. 2: Geodesic in a medium with dispiration,  $\alpha=1/10$, $\beta\neq 0$.
\\
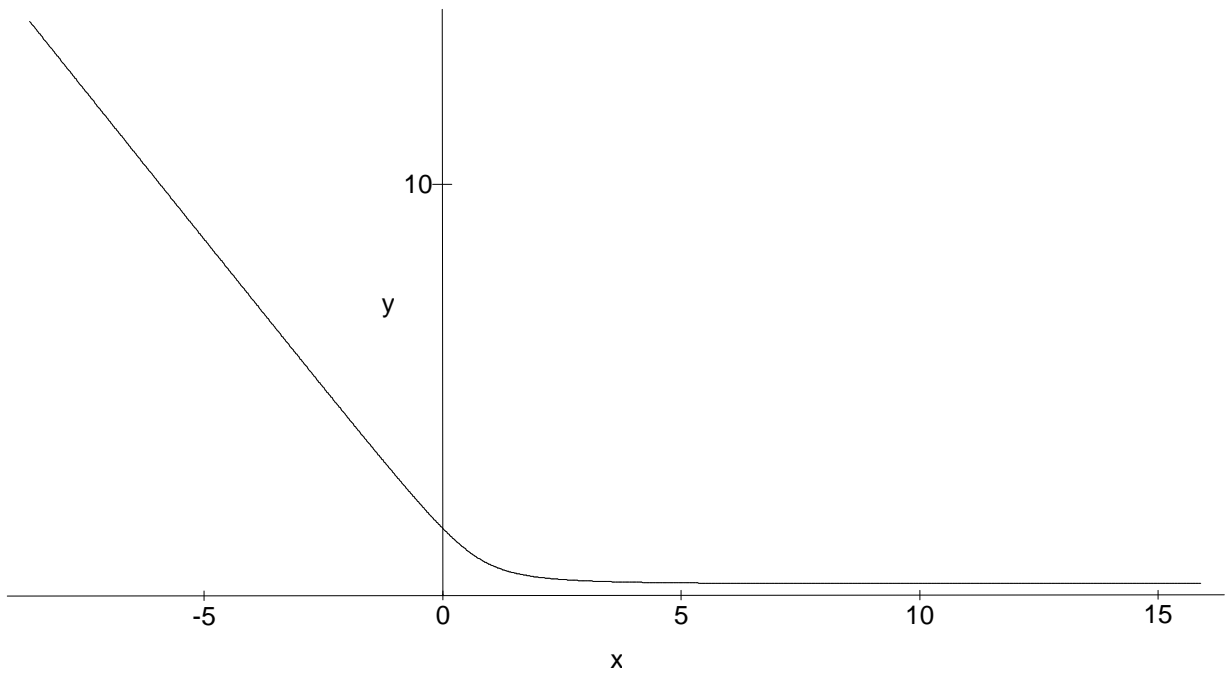
Fig. 3: Geodesic in a disclinated medium, $\alpha=7/5$, $\beta=0$.
\\
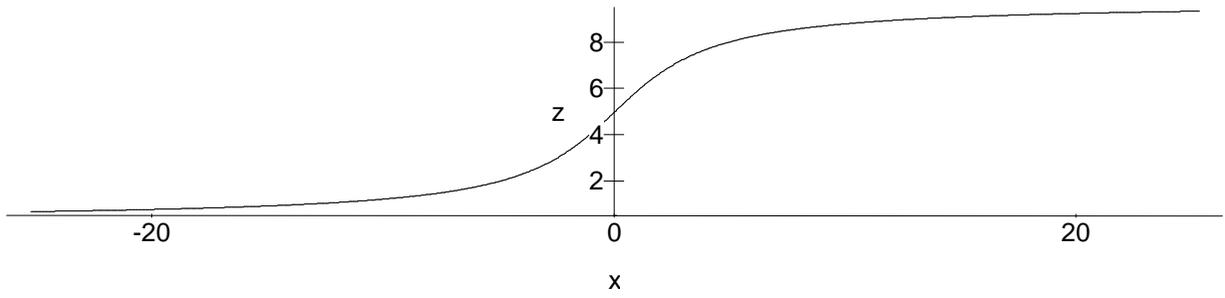
Fig. 4: Geodesic in a medium with a screw dislocation, $\alpha=1$, $\beta\neq 0$.

\end{document}